\UseRawInputEncoding

\documentclass[aip,pre,amsmath,amssymb,floatfix,reprint,citeautoscript,noeprint,superscriptaddress,twocolumn]{revtex4-2}

\usepackage{dsfont}
\usepackage{lipsum} 
\usepackage{bibentry}
\usepackage[english]{babel}
\usepackage[dvipsnames]{xcolor}
\usepackage{graphicx}
\usepackage[caption=false]{subfig} 
\usepackage{amsmath,amssymb,bm}
\usepackage[version=3]{mhchem}
\usepackage{verbatim}
\usepackage{multirow}
\usepackage{dcolumn}
\usepackage{float}
\usepackage{nicefrac}
\usepackage{siunitx}
\usepackage{booktabs}
\usepackage{chemformula}
\usepackage{wrapfig}
\usepackage{enumitem}  
\usepackage{transparent}
\usepackage[colorlinks,allcolors=black,citecolor=blue,urlcolor=blue]{hyperref}
\usepackage{tcolorbox}
\usepackage{relsize}

\hypersetup{
    citecolor=Turquoise,
    colorlinks=true,
    linkcolor=black,    
    urlcolor=Turquoise,
    }
\DeclareSIUnit[number-unit-product = {\,}]{\amu}{amu}
\DeclareSIUnit[number-unit-product = {\,}]{\kJmol}{\kilo\joule\per\mol}
\DeclareSIUnit[number-unit-product = {\,}]{\Nsm}{\newton\second\per\meter\cubed}
\DeclareSIUnit[number-unit-product = {\,}]{\THz}{\tera\hertz}
\DeclareSIUnit[number-unit-product = {\,}]{\meV}{\milli\electronvolt}
\DeclareSIUnit[number-unit-product = {\,}]{\cal}{cal}

\newcommand{\mbf}[1]{\boldsymbol{\mathit{#1}}}
\newcommand{\mrm}[1]{\mathrm{#1}}
\newcommand{\mcl}[1]{\mathcal{#1}}
\newcommand{\tcr}[1]{\textcolor{black}{#1}}

\newcommand{\etal}{\emph{et al.}}

\usepackage{sansmathfonts}
\usepackage[justification=centerlast, font={sf}, labelfont={bf}]{caption}
\begin{document}

\title{The Roles of Bulk and Surface Thermodynamics in the
    Selective Adsorption of a Confined Azeotropic Mixture}

\author{Katie L. Y. Zhou}

\affiliation{Department of Chemistry, Durham University, South Road, Durham, DH1
3LE, United Kingdom}

\author{Anna T. Bui}

\affiliation{Yusuf Hamied Department of Chemistry, University of
  Cambridge, Lensfield Road, Cambridge, CB2 1EW, United Kingdom}
\affiliation{Department of Chemistry, Durham University, South Road, Durham, DH1
3LE, United Kingdom}

\author{Stephen J. Cox}
\email{stephen.j.cox@durham.ac.uk}

\affiliation{Department of Chemistry, Durham University, South Road, Durham, DH1
3LE, United Kingdom}

\date{\today}

\color{darkgray}
%\pagecolor{lightgray}

\begin{abstract}
  \textbf{Abstract:} Fluid mixtures that exhibit an azeotrope cannot be purified by
  simple bulk distillation. Consequently, there is strong motivation
  to understand the behavior of azeotropic mixtures under
  confinement. We address this problem using a
  machine-learning-enhanced classical density functional theory
  \tcr{(cDFT)} applied to a binary Lennard--Jones mixture that
  exhibits azeotropic phase behavior. As proof-of-principle of a
  ``train once, learn many'' strategy, our approach combines a neural
  functional trained on a single-component repulsive reference system
  with a mean-field treatment of attractive interactions,
  \tcr{inspired by the connection between cDFT and local molecular
    field theory}. The theory faithfully describes capillary
  condensation and results from grand canonical Monte Carlo
  simulations. Moreover, by taking advantage of a known accurate
  equation of state, the \tcr{``neural LMFT''} we present
  well-describes bulk thermodynamics by construction. Exploiting the
  computational efficiency of \tcr{neural LMFT}, we systematically
  evaluate adsorption selectivity across a wide range of compositions,
  pressures, temperatures, and wall--fluid affinities. In cases where
  the wall--fluid interaction is the same for both species, we find
  that the pore becomes completely unselective at the bulk azeotropic
  composition. Strikingly, this unselective point persists far from
  liquid--vapor coexistence, including in the supercritical
  regime. Analysis of the bulk equation of state across a wide range
  of thermodynamic state points shows that the azeotropic composition
  coincides with equal partial molar volumes and an extremum in the
  isothermal compressibility. A complementary thermodynamic analysis
  demonstrates that unselective adsorption corresponds to an aneotrope
  (a point of zero relative adsorption) and an extremum in the
  interfacial free energy. We also find that the two interfaces of the
  slit pore behave independently down to remarkably small slits.
\end{abstract}

\maketitle

%\color{darkgray}

\section{Introduction}

A detailed understanding of the thermophysical properties of fluid
mixtures under confinement is crucial for a wide range of
applications, including industrial separation processes, gas storage,
and healthcare \cite{shollSevenChemicalSeparations2016,
  furukawaChemistryApplicationsMetalOrganic2013,
  capelo-avilesSelectiveAdsorptionCO22025}. Specific examples span oil
recovery \cite{barsottiReviewCapillaryCondensation2016}, hydrogen
storage \cite{clauzierEnhancedH2Uptake2012,
  clauzierEnhancedH2Uptake2014}, and microfluidic systems
\cite{yangCapillaryCondensationAtomicscale2020,
  khanjaniCapillaryMicrofluidicsDiagnostic2025}. From a theoretical
standpoint, molecular simulations are widely regarded as the method of
choice \cite{wildingCriticalpointCoexistencecurveProperties1995,
  hoSolubilityGasesWater2015, huEffectConfinementNanoporous2016,
  Fertig2024, wangIntegratingMolecularDynamics2024,Shen2005}, offering
microscopic insight that is often inaccessible to experiment alone
\cite{kellayPrewettingBinaryLiquid1993,
  bonnWettingLayeringCritical1993}. However, traditional approaches
rooted in liquid state theory are experiencing a renaissance
\cite{buiClassicalDensityFunctional2024,
  wuPerfectingLiquidStateTheories2023}. In particular, classical
density functional theory (cDFT) is attracting renewed interest in
physical chemistry, driven by the ability of machine learning (ML) to
deliver highly accurate and computationally efficient approximations
\cite{sammullerNeuralFunctionalTheory2023,
  buiLearningClassicalDensity2025,
  robitschkoLearningBulkInterfacial2025,
  sammullerNeuralDensityFunctional2025,
buiDielectrocapillarityExquisiteControl2025,ahmedPhaseDiagramWeeksChandlerAndersen2009,catsMachinelearningFreeenergyFunctionals2021,
  yangHighDimensionalOperatorLearning2024,
  linAnalyticalClassicalDensity2020}.

In this article, our aim is to demonstrate a recently proposed
ML-based approach to cDFT for a simple Lennard--Jones (LJ) fluid
mixture. While similar ideas have been explored recently in
Ref.~\onlinecite{robitschkoLearningBulkInterfacial2025}, our approach
differs in two key respects. First, methodologically, we train the ML
model on a repulsive \emph{single-component} reference system and
incorporate attractive interactions through a simple mean-field
treatment---an approach that enhances transferability among
systems. Second, from an application standpoint, we focus on a binary
mixture with asymmetric interactions, which exhibits azeotropic phase
behavior. As we will see, the existence of an azeotrope appears to
strongly influence the fluid's behavior---both in bulk and under
confinement---across a broad range of thermodynamic
conditions.

The potential appeal of cDFT over molecular simulations becomes
immediately evident when considering its fundamental formalism \cite{Evans1979,
hansenTheorySimpleLiquids2013}. The
central object in cDFT is the grand potential functional
\begin{multline}
  \label{eqn:varOmega}
\varOmega_{V}\big([\{\varrho_\alpha\}], T\big) = \mathcal{F}^{(\rm id)}_{\text{intr}}\big([\{\varrho_\alpha\}], T\big)
    + \mathcal{F}^{(\rm ex)}_{\text{intr}}\big([\{\varrho_\alpha\}], T\big)\\
    + \sum_\alpha \int\!\text{d}\mbf{r}\,\varrho_\alpha(\mbf{r})\big[ V_{\alpha}(\mbf{r}) - \mu_\alpha \big],
\end{multline}
where $\mathcal{F}^{(\rm id)}_{\text{intr}}$ and
$\mathcal{F}^{(\rm ex)}_{\text{intr}}$ are, respectively, the ideal
and excess intrinsic Helmholtz free energy functionals\tcr{, and $T$
  is the temperature}. A particle of species $\alpha$, with chemical
potential $\mu_\alpha$, experiences an external one-body potential
$V_{\alpha}$, while $\varrho_{\alpha}$ denotes its average one-body
density, though not necessarily at equilibrium. Functional
minimization of $\varOmega_{V}$ with respect to $\varrho_\alpha$
yields the equilibrium one-body density of species $\alpha$,
$\rho_\alpha$, which satisfies the Euler--Lagrange equation,
\begin{equation}
   \label{eqn:EL}
  \Lambda^3_\alpha \rho_{\alpha}(\mbf{r}) = \exp\left(-\beta\big[V_{\alpha}(\mbf{r})-\mu_\alpha\big]+ c^{(1)}_\alpha(\mbf{r};[\{\rho_\alpha\}],T)\right)
\end{equation}
with $\Lambda_{\alpha}$ denoting the thermal de Broglie wavelength of
species $\alpha$, $\beta = 1/k_{\rm B}T$ ($k_{\rm B}$ is the Boltzmann
constant), and
\begin{equation}
  \label{eqn:c1-derivative}
  c^{(1)}_\alpha(\mbf{r};[\{\varrho_\alpha\}], T) = -\frac{\delta\beta\mathcal{F}^{(\rm ex)}_{\rm intr}([\{\varrho_\alpha\}], T)}{\delta\varrho_\alpha(\mbf{r})},
\end{equation}
is the one-body direct correlation functional. In the context of
the Euler--Lagrange equation, $c^{(1)}_\alpha$ acts as a
self-consistent one-body potential that accounts for the effects of
correlations on the structure of the fluid. With the equilibrium
densities from Eq.~\ref{eqn:EL}, the grand potential of the system
readily follows: $\Omega = \varOmega_{V}([\{\rho_\alpha\}],T)$.  The
procedure for obtaining the structure and thermodynamics of the system
can, therefore, be succinctly put: minimize a functional to obtain the
equilibrium structure, then evaluate the functional at equilibrium to
obtain the thermodynamic potential of the system. This is a far
simpler operation than explicitly sampling the many-body equilibrium
distribution function with molecular simulations.

What, then, has prevented cDFT from becoming the method of choice for
understanding the equilibrium properties of fluids? The answer is
straightforward. While a rigorous theoretical framework, in practice
cDFT relies upon approximations to
$\mathcal{F^{(\rm ex)}_{\rm intr}}$. (Only for hard rods in a single
dimension is $\mathcal{F^{(\rm ex)}_{\rm intr}}$ known exactly
\cite{percus1976equilibrium}.) In the case of hard spheres, very good
approximations founded on Rosenfeld's fundamental measure theory (FMT)
have been available for several decades
\cite{rosenfeldFreeenergyModelInhomogeneous1989,
  rothFundamentalMeasureTheory2010, rothFundamentalMeasureTheory2002,
  hansen-goosDensityFunctionalTheory2006}. When combined with an
\emph{ad hoc} mean-field treatment of attractive interactions, cDFT
has been shown to be a powerful practical method for understanding the
physics of simple liquids. In particular, the physics of capillarity
\cite{evansFluidsNarrowPores1986a, evansPhaseEquilibriaSolvation1987},
surface drying and wetting \cite{evansCriticalDryingLiquids2016,
  evansUnifiedDescriptionHydrophilic2019}, and solvophobicity
\cite{Coe2022} have been extensively studied in single-component
systems, owing to the ability of the mean-field approximation to
qualitatively capture liquid--gas phase coexistence. While less
intensely studied compared to single-component fluids, such a
mean-field approach has also provided qualitative insight into the
behavior of simple mixtures \cite{tanSelectiveAdsorptionSimple1992,
  kierlikDensityfunctionalTheoryInhomogeneous1991,
  kierlikLiquidLiquidEquilibrium1995,
  taghizadehPopulationInversionBinary2011,
  jiangTheoryAdsorptionTrace1994, qiaoEnhancingGasSolubility2020,
  Denton1990, Napari1999}, though the richer phase behavior of
multicomponent fluids---including liquid--liquid coexistence,
azeotropy and heteroazeotropy \cite{vanKonynenburg1980}---pushes the
limits of analytical approximations for
$\mathcal{F}^{(\mathrm{ex})}_{\mathrm{intr}}$.

Going beyond hard spheres and a mean-field treatment for attractive
interactions \tcr{is challenging, though we note significant progress
  with analytical approaches has been made}
\cite{tang2003density,tang2004modeling,fu2004self,sauer2017classical}. Naturally,
several groups have recently turned to ML. For example, early works by
Oettel and co-workers focused on one-dimensional systems
\cite{linAnalyticalClassicalDensity2020,
  shang-chunClassicalDensityFunctional2019} and anisotropic patchy
particles \cite{Simon2024}, while Cats \etal{} used ML to improve the
standard mean-field approximation for the three-dimensional LJ fluid
\cite{catsMachinelearningFreeenergyFunctionals2021}.  One ML scheme
that is gaining significant traction is ``neural functional theory''
\cite{sammullerNeuralFunctionalTheory2023} introduced by Samm\"{u}ller
\etal{} The basis of this physics-informed approach is the
Euler--Lagrange equation (Eq.~\ref{eqn:EL}). For known
$\{V_{\alpha}\}$ and $\{\mu_\alpha\}$, with $\{\rho_\alpha\}$ obtained
from grand canonical Monte Carlo (GCMC) simulation, \tcr{the spatial
  variation of each $c^{(1)}_\alpha$ at equilibrium is} determined by
rearranging Eq.~\ref{eqn:EL}:
\begin{equation}
  \label{eqn:central-ML}
  c^{(1)}_\alpha(\mbf{r};[\{\rho_\alpha\}],T) = \ln\Lambda_\alpha^3\rho_\alpha(\mbf{r}) + \beta\big(V_{\alpha}(\mbf{r})-\mu_\alpha\big).
\end{equation}
By obtaining $\{c^{(1)}_\alpha\}$ from \tcr{many GCMC} simulations
with different \tcr{$\{V_{\alpha}\}$, $\{\mu_\alpha\}$, and
  equilibrium density profiles $\{\rho_\alpha\}$,} a training set is
established to learn the local functional dependence of
$c^{(1)}_\alpha$ on $\{\varrho_\alpha\}$ with a neural network.

Originally developed for the single-component hard-sphere
fluid---where it even outperformed FMT-based functionals
\cite{sammullerNeuralFunctionalTheory2023}---the neural cDFT approach
has since been extended to more complex systems. For example,
Samm\"{u}ller \etal{} applied neural cDFT to the LJ fluid,
demonstrating accurate predictions of liquid--vapor coexistence
\cite{sammullerNeuralDensityFunctional2025}.  Building on this,
Robitschko \etal{} showed that liquid--liquid coexistence in a simple
binary LJ mixture is also well captured
\cite{robitschkoLearningBulkInterfacial2025}.  By incorporating
orientational correlations, Yang \etal{} have successfully applied
neural cDFT to molecular \ce{CO2}
\cite{yangHighDimensionalOperatorLearning2024}.

While neural cDFT exploits the local nature of correlations, Bui and
Cox addressed systems with long-ranged electrostatic interactions
\cite{buiLearningClassicalDensity2025}, combining neural cDFT with a systematically
improvable mean-field approach inspired by local molecular field
theory (LMFT) \cite{rodgersLocalMolecularField2008}. Initially
developed for primitive electrolyte models
\cite{buiLearningClassicalDensity2025}, this LMFT-style framework was
later generalized to systems with nontrivial coupling between charge
and number density, such as polar fluids
\cite{buiFirstPrinciplesApproach2025,
  buiDielectrocapillarityExquisiteControl2025}.  At the foundation of
this generalization lies hyperdensity functional theory
\cite{sammullerHyperdensityFunctionalTheory2024} (hyper-DFT), also
introduced by Samm\"{u}ller \etal{} in the context of soft matter,
which establishes that any equilibrium observable can be expressed as
a functional of the one-body density. Leveraging hyper-DFT, Bui and
Cox established a rigorous framework for electromechanics in fluids
\cite{buiFirstPrinciplesApproach2025} and demonstrated that electric
field gradients can alter liquid--vapor coexistence in polar fluids,
including water
\cite{buiDielectrocapillarityExquisiteControl2025}. This work
uncovered a previously unknown effect---dielectrocapillarity---in
which field gradients control adsorption into porous media.

Clearly, the neural functional approach offers a route to fully
harness the advantages of cDFT over molecular simulations. While the
\tcr{LMFT-based approach} arguably remains essential for systems
dominated by long-range interactions, many systems can now be tackled
simply by learning $\{c^{(1)}_{\alpha}\}$, without invoking any of the
standard approximations of liquid state theory. In this light, can
traditional strategies---such as decomposing the system into a
repulsive reference plus mean-field attraction, as in FMT-based
approaches---still play a meaningful role?

We argue that the answer is ``yes.'' Our reasons are twofold. First,
when the bulk equation of state (EoS) is well established by other
means, the \tcr{LMFT-based} framework described in
Ref.~\onlinecite{buiFirstPrinciplesApproach2025} offers a
straightforward way to leverage the rich information that it
encodes. By adopting this strategy, we can focus efforts where they
are most needed: neural cDFT can be targeted to the study of
inhomogeneous systems where its advantages are most pronounced,
leaving the bulk physics to the already-known EoS.  Second, while cDFT
itself is more efficient than molecular simulations, neural cDFT comes
with an initial computational overhead in obtaining the training
data. Decomposing the system into a repulsive reference---treated with
neural cDFT---and mean-field attractive interactions, opens the door
to a ``train once, learn many'' strategy, where a single, accurate
repulsive reference can be reused across multiple systems. Such an
approach becomes particularly advantageous in the case of mixtures; in
this article, we demonstrate a proof-of-principle for a binary LJ
mixture with asymmetric interactions that exhibits azeotropy.

The rest of the article is organized as follows. In Sec. \tcr{II.A}, we
introduce the LJ mixture and outline the underpinning \tcr{LMFT-based}
framework. \tcr{Sections II.B--II.D provide details on the neural cDFT setup.} In Secs. \tcr{III.A and III.B}, we validate the approach against GCMC
simulations and examine adsorption and selectivity in a slit-pore that
interacts with both species in the same manner. We analyze the bulk
thermodynamic behavior of the mixture, with particular emphasis on the
role of the azeotropic composition across a wide range of pressures
and temperatures, in Sec. \tcr{III.C}. In Sec. \tcr{III.D}, we develop a thermodynamic
description of pore selectivity that sheds light on the underlying
driving forces for selective adsorption.
%
%clarifies the interplay between bulk and interfacial contributions,
%including the effects of selective wall--fluid interactions.
%
Finally, Sec. \tcr{IV} summarizes our main findings and discusses their
broader implications and possible extensions.

\begin{figure*}[t!]
  \centering
  \includegraphics{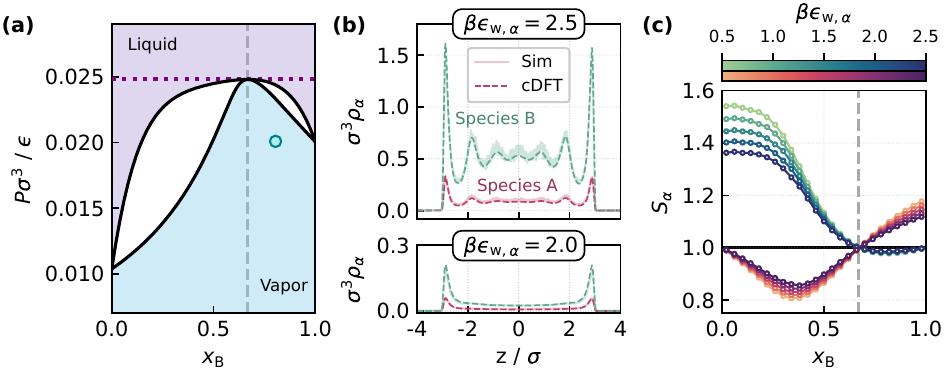}
  \caption{Effects of confinement on a binary LJ mixture. (a) The
    pressure-composition phase diagram of the mixture at
    $k_{\rm B}T/\epsilon = 0.77 $. An azeotrope forms at
    $x_{\rm B}^{\rm{(az)}} \approx 0.67$, as indicated by the vertical
    dashed line. \tcr{(b) Density profiles for the mixture confined in
      a slit pore, in equilibrium with a reservoir at
      $P\sigma^3/\epsilon = 0.020$, and $x_{\rm B} = 0.78$ [indicated
      by the circle in (a)]; we consider two different wall--fluid
      interaction strengths, as indicated by the labels.} Good
    agreement between the theory and the simulation data is
    observed. (c) Pore selectivities vs $x_{\rm B}$\tcr{, at the
      azeotropic pressure [indicated by the dotted line in (a)], for
      different wall--fluid interactions. Results are shown for both}
    component A (orange to purple) and B (green to blue). Lines serve
    as a guide to the eye. A reversal in selectivity is observed at
    $x_{\rm B}^{\text{(az)}}$, as indicated by the vertical dashed
    line.}
\label{fig:1}
\end{figure*}

\section{Methods}
\subsection{Neural LMFT}

The intermolecular interactions that govern fluid mixtures can, in
general, be complicated, potentially comprising strongly directional
interactions in the form of hydrogen bonds \cite{Noskov2005},
long-ranged electrostatic effects \cite{Dhattarwal2025, Joung2008},
and even metallicity \cite{Rosenbrock2021}.  Nonetheless, many of the
most salient aspects of fluid mixtures are captured by relatively
simple forms for the intermolecular interactions.  For instance,
complex liquid--vapor and liquid--liquid phase equilibria are
reproduced with LJ mixtures
\cite{stephanInterfacialPropertiesBinary2019,
  stephanInfluenceDispersiveLongrange2020}, alkane phase separation
with generalized Mie potentials \cite{Potoff2009}, binary alloys
with Stillinger--Weber \cite{Laradji1995, Stillinger1985} and Tersoff
\cite{Tersoff1986, Kelires1989} potentials, and polymer demixing with
Gaussian core models \cite{Archer2001}.

Here, the system we investigate is a binary mixture of species A and
B, whose pairwise interactions are prescribed by the truncated and
shifted LJ pair potential,
\begin{equation}
  \tcr{u_{\alpha\eta}(r) = 4\epsilon_{\alpha\eta}[(\sigma/r)^{12} - (\sigma/r)^{6} ] - 4\epsilon_{\alpha\eta}[(\sigma/r_{\mrm{c}})^{12} - (\sigma/r_{\rm c})^{6}]}
\end{equation}
\tcr{for $r<r_{\rm c}$ (and zero otherwise)}, where $\alpha$ and
$\eta$ are species labels, $r_{\rm c} = 2.5\sigma$, and the molecular
diameter, $\sigma$, is the same for all particles. In contrast, the
strength of interaction between particles is species-dependent.
Specifically, $\epsilon_{\rm AA} = \epsilon$,
$\epsilon_{\rm BB} = 0.9\epsilon$, and
$\epsilon_{\rm AB} = 0.806\epsilon$. In Fig.~\ref{fig:1}a we present
the predicted bulk phase diagram from the PeTS EoS
\cite{heierEquationStateLennardJones2018,
  rehnerFeOsOpenSourceFramework2023} at a temperature
$k_{\rm B}T/\epsilon = 0.77$, in the $P$--$x_{\rm B}$ plane, where $P$
is pressure and $x_{\rm B}$ is the mole fraction of B. Previous
studies \cite{stephanInterfacialPropertiesBinary2019,
  stephanInfluenceDispersiveLongrange2020,
  stephanMolecularInteractionsVaporliquid2020} have shown that, at
this temperature, this system exhibits a positive azeotrope---the
point on the phase diagram at which liquid and vapor have the same
composition---at mole fraction $x_{\rm B}^{\rm (az)} \approx 0.67$ and
pressure $P^{\rm (az)}\sigma^3/\epsilon = 0.0248$.

To describe the structure and thermodynamics of this system in a
density functional framework, we will adopt \tcr{an LMFT-based
  approach to neural cDFT, which we refer to as ``neural LMFT'' for
  convenience. In LMFT, as developed by Weeks and co-workers
  \cite{weeksRolesRepulsiveAttractive1998,
    weeksSelfConsistentTreatmentRepulsive1995,
    remsingLongrangedContributionsSolvation2016}, a short-ranged
  reference system with pairwise interactions $u_{0,\alpha\eta}(r)$ is
  introduced.  For this reference system, we then seek the set of
  one-body potentials $\{\phi_{{\rm R},\alpha}\}$ such that the
  resulting equilibrium one-body densities satisfy
  $\rho_{{\rm R},\alpha}(\boldsymbol{r}) =
  \rho_\alpha(\boldsymbol{r})$ for all $\alpha$, where the subscript
  `R' indicates properties pertaining to the reference system.}

\tcr{The formal similarity between LMFT and cDFT in a mean-field
  approximation was originally established by Archer and Evans
  \cite{archerRelationshipLocalMolecular2013}, though the
  well-controlled nature of the mean-field form that underlies LMFT
  was left implicit
  \cite{remsingLongrangedContributionsSolvation2016}. Building on
  Ref.~\onlinecite{archerRelationshipLocalMolecular2013}, LMFT's
  equivalence to hyper-DFT was also recently formalized
  \cite{buiFirstPrinciplesApproach2025}, re-establishing the
  well-controlled nature of the mean-field approximation.}

\tcr{More important for the present study, however, is that
  Ref.~\onlinecite{buiFirstPrinciplesApproach2025} asserted a
  mean-field free energy functional in which the bulk and
  inhomogeneous contributions are cleanly separated,}
\begin{align}
\label{eqn:mean-field}
\mathcal{F}^{\rm (ex)}_{\rm intr}&(\tcr{[\{\varrho_\alpha\}]},T)
= \mathcal{F}^{\rm (ex)}_{\rm intr,R}(\tcr{[\{\varrho_\alpha\}]},T)
+ \sum_\alpha \Delta\mu_\alpha\!
  \int\!\! \mathrm{d}\mbf{r}\,\varrho_\alpha(\mbf{r})
\nonumber\\
& + \frac12 \sum_{\alpha,\eta}
\int\!\!\mathrm{d}\mbf{r}
\!\!\int\!\! \mathrm{d}\mbf{r}'\,
\delta_{\rm u}\tcr{\varrho_{\alpha}}(\mbf{r})
u_{1,\alpha\eta}(|\mbf{r}-\mbf{r}'|)
\,\delta_{\rm u}\tcr{\varrho_{\eta}}(\mbf{r}').
\end{align}
where $\Delta\mu_\alpha = \mu_\alpha-\mu_{\rm R,\alpha}$ is the
difference in chemical potentials between the LJ and reference
systems, $u_{1,\alpha\eta} = u_{\alpha\eta} - u_{0,\alpha\eta}$, and
\tcr{$\delta_{\mrm{u}}\varrho_{\alpha} = \varrho_{\alpha} - \bar{\rho}_{\alpha}$}, with
$\bar{\rho}_{\alpha}$ denoting the uniform bulk density for
species $\alpha$ \tcr{for given  
$\{\mu_\alpha\}$ and $T$.} 

\tcr{With such a mean-field form, following either
  Ref.~\onlinecite{buiFirstPrinciplesApproach2025} or
  Ref.~\onlinecite{archerRelationshipLocalMolecular2013},} one can
show that
\begin{equation}
  \begin{split}
    c^{(1)}_\alpha(\mbf{r};[\tcr{\{\rho_{\alpha}\}]}, T) =  c^{(1)}_{{\rm R},\alpha}(\mbf{r};\tcr{[\{\rho_{\alpha}\}]}, T) - \beta\phi_{\mrm{R},\alpha}(\mbf{r}) -\beta\Delta\mu_\alpha,
  \end{split}
  \label{eqn:c1}
\end{equation}
where
\begin{equation}
  \begin{split}
    \label{eqn:phiR}
    \phi_{\mrm{R},\alpha}(\mbf{r}) =
    \sum_{\eta}\int\!\mrm{d}\mbf{r}^\prime\,u_{1,\alpha\eta}(|\mbf{r}-\mbf{r}^\prime|) \delta_{\mrm{u}}\tcr{\rho_{\eta}}(\mbf{r}^\prime).
  \end{split}
\end{equation}
\tcr{The separation of bulk and inhomogeneous contributions to the
  mean-field free energy functional} affords a level of flexibility
that can be put to practical advantage; an accurate description of the
bulk fluid can be obtained if $\{\Delta\mu_\alpha\}$ is known by any
means. For example, in the case of primitive models of electrolytes
and dielectric fluids, analytic expressions for $\{\Delta\mu_\alpha\}$
have been derived based on Stillinger--Lovett sum rules
\cite{buiFirstPrinciplesApproach2025,
  buiLearningClassicalDensity2025}. Here, we will take advantage of
the fact that $\{\Delta\mu_\alpha\}$ can be obtained from a known EoS.

Success of this \tcr{neural LMFT} approach rests upon a suitable
choice of reference system. For the binary LJ mixture that we study
here, in which A and B have the same molecular diameter, we choose the
reference system to be the purely repulsive \emph{single-component}
fluid described by the Weeks--Chandler--Anderson (WCA) potential
\cite{weeksRoleRepulsiveForces1971}, i.e., $u_{0,\alpha\eta}=u_0$, with
\begin{equation}
  u_{0}(r) = \begin{cases}
    4\epsilon[(\sigma/r)^{12} - (\sigma/r)^{6}] + \epsilon, & r<2^{1/6}\sigma \\
    0, &  r \ge 2^{1/6}\sigma.
  \end{cases}
\end{equation}
Combining Eqs.~\ref{eqn:EL} and~\ref{eqn:c1}, the Euler--Lagrange
equation for species $\alpha$ of the LJ mixture reads,
%
%\begin{equation}
  \begin{align}
    \label{eqn:EL2}
    &\Lambda^{3}_\alpha\rho_{\alpha}(\mbf{r}) = \\
    &\exp\big(-\beta\big[V_{\alpha}(\mbf{r}) + \phi_{\mrm{R},\alpha}(\mbf{r}) - \mu_{\mrm{R},\alpha} \big] 
     + c^{(1)}_{\mrm{R}}(\mbf{r};[\rho_\mrm{A} + \rho_\mrm{B}],T)\big), \nonumber
  \end{align}
%\end{equation}
%
with
\begin{equation}
 \label{eqn:mu_R}
  \mu_{\rm R, \alpha}(\bar{\rho}_{\rm A},\bar{\rho}_{\rm B}) =
  k_{\rm B}T\ln\Lambda_{\alpha}^3\bar{\rho}_{\alpha} - k_{\rm B}Tc^{(1)}_{\rm R}([\bar{\rho}_{\rm A}+\bar{\rho}_{\rm B}],T).
\end{equation}

\subsection{Generation of training data} 

Following
Ref.~\onlinecite{sammullerNeuralFunctionalTheory2023},  we trained a neural functional to
represent $c^{(1)}_{\mrm{R}}$ of the WCA fluid, using data from 900 GCMC
simulations.  In practice,
the neural functional is limited to a planar geometry, i.e.,
$c^{(1)}_{\mrm{R}}(\mbf{r},[\{\varrho_\alpha\}], \tcr{T})\to
c^{(1)}_{\mrm{R}}(z,[\{\varrho_\alpha\}], \tcr{T})$; we direct the reader
toward Ref.~\onlinecite{glitsch2025neural} for recent developments
that extend the neural functional approach to resolution in higher
dimensions.

Simulations were performed using our own code available on Github
(\url{https://github.com/annatbui/gcmc}) or Zenodo \cite{gcmc}. They
were conducted at temperatures $k_{\rm B}T/\epsilon=1.0$, $1.5$
and~$2.0$ with randomized chemical and external potentials. The
chemical potentials were chosen from the range
$-10 \leq \beta \mu \leq 3$ and the external potentials had the form
\begin{equation}
   V(z) =
    \sum^4_{n=1} A_n \sin \left( \frac{2\pi nz}{\ell} + \Phi_n \right) + \sum_n V^{\text{lin}}_n (z)
\end{equation}
where $A_n$ were randomly chosen Fourier coefficients from a normal
distribution of variance $2.5(k_{\rm B}T)^2$ and the phases $\Phi_n$
were chosen uniformly between 0 and $2\pi$.  The simulation box was
cubic with length $\ell = 10\sigma$ and periodic boundary conditions
were applied.  The linear function $V_n^{\text{lin}}(z)$ takes the
form
\begin{equation}
V_n^{\text{lin}}(z) = V_{n,1} + \frac{(V_{n,2} - V_{n,1})(z - z_{n,1})}{(z_{n,2} - z_{n,1})}
\end{equation}
for $z_{n,1} < z < z_{n,2}$ and 0 otherwise, where $z_{n,1}$ and
$z_{n,2}$ were uniformly chosen such that
$ 0 < z_{n,1} < z_{n,2} < \ell$, and $V_{n,1}$ and $V_{n,2}$ were
randomly chosen from a normal distribution with variance
$4(k_{\rm B}T)^2$. Each external potential had four sinusoidal
segments, and between $1$ and $5$ linear segments. Half of the
potentials had planar hard walls, where $V(z) = \infty$ for
$z \leq z_{\text{w}}/2$ and $z \geq \ell - z_{\text{w}}/2$;
$z_{\text{w}}$ was randomly chosen uniformly between $1\sigma$ and
$3\sigma$.

Each simulation was run for $10^9$ steps, with equilibration for
$10^6$ steps. For each simulation, the planar density profile
$\rho(z)$ was obtained from a histogram of the positions of the
particles, and $c^{(1)}_{\mrm{R}}(z;[\rho],T)$ was obtained from
numerical evaluation of the single component \tcr{version} of
Eq. \ref{eqn:central-ML}. The total computation time for the
generation of the entire training dataset is on the order of $10^4$
CPU hours.

\subsection{Training the neural functional}
Our training procedure
largely follows that of previous work
\cite{sammullerNeuralFunctionalTheory2023,
  sammullerNeuralDensityFunctional2025,
  robitschkoLearningBulkInterfacial2025,
  buiLearningClassicalDensity2025}. A neural network was trained to
represent $c^{(1)}_{\mrm{R}}(z;[\rho],T)$, implemented using
Keras/Tensorflow with the standard Adam optimizer
\cite{kingmaAdamMethodStochastic2015}. The input layers take in the
temperature and the density in a window of size $3\sigma$ around the
location of interest, with spatial discretization
$\Delta z = 0.005\sigma$. This was followed by two hidden layers of 32
nodes with softplus activation functions, and then the single-node
output. The simulation dataset was split 3:1:1 for the training,
validation, and test datasets, respectively. Data augmentation allowed
the doubling of the training data through mirroring. The model was
trained for 100 epochs in batches of size 256 with the mean squared
error as the loss function. The initial learning rate was 0.001,
decreasing by 5\% per epoch. The training was done on an
\texttt{NVIDIA GH200 Grace Hopper Superchip} in under an hour\tcr{; we
  stress that use of this architecture was made out of convenience due to
  available resources, and training is entirely possible on commodity
  hardware} \cite{sammullerNeuralFunctionalTheory2023,
  sammullerNeuralDensityFunctional2025,
  robitschkoLearningBulkInterfacial2025,
  buiLearningClassicalDensity2025,
  buiDielectrocapillarityExquisiteControl2025}.

\subsection{Using the neural functional}

Once the neural functional has been trained, combining with LMFT using
Eq.~\ref{eqn:c1} gives
\tcr{$c_\alpha^{(1)}(z;[\{\rho_\alpha\}],T)$}. Given a bulk state
point, $\mu_{\alpha}(\bar{\rho}_{\rm A},\bar{\rho}_{\rm B})$ can be
obtained directly from the PeTS EoS
\cite{heierEquationStateLennardJones2018} \tcr{(here we use its
  implementation from the FeOs package
  \cite{rehnerFeOsOpenSourceFramework2023} version 0.8.0)}\tcr{;
  $\Delta\mu_\alpha$ is then obtained by subtracting
  $\mu_{\rm R, \alpha}(\bar{\rho}_{\rm A},\bar{\rho}_{\rm B})$ given
  by Eq.~\ref{eqn:mu_R}.} Minimization of the Euler--Lagrange equation
to obtain density profiles is done self-consistently using a mixed
Picard scheme, and typically takes around 2 minutes on a standard CPU
(and faster on a GPU).

To evaluate the grand potential of the binary LJ mixture, we use
Eq.~\ref{eqn:varOmega} with
$\mathcal{F}^{\mrm{(ex)}}_{\mrm{intr}}$ obtained by functional line
integration \cite{sammullerNeuralFunctionalTheory2023},
\begin{equation}
\beta\mathcal{F}^{(\rm ex)}_{\text{intr}}/\mcl{A} = -\int^1_0 \!\!\mrm{d}\lambda\,\sum_{\alpha}\int\!\!\mrm{d} 
z \, \rho_{\alpha}(z) c^{(1)}_{\alpha}(z; [\{\lambda\rho_{\alpha}\}], T),
\end{equation}
\tcr{where $\mathcal{A}$ is the cross-sectional area of the slit pore, }
and $\beta\mathcal{F}_{\text{intr}}^{\text{(id)}}/\mcl{A} = \sum_\alpha
\int \text{d} z \, \rho_\alpha (z)\left( \ln[\Lambda_\alpha^3
  \rho_\alpha (z)] - 1 \right) $. Overall we have performed approximately 6200 cDFT
calculations, from which we directly obtain both equilibrium density
profiles and free energies.

\section{Results and Discussion}
\label{sec:results}

\tcr{\subsection{Neural LMFT Successfully Describes an Asymmetric Binary LJ Mixture}}

In Fig.~\ref{fig:1}b we show results from \tcr{neural LMFT}
for the binary LJ mixture in a slit-pore at equilibrium with a
reservoir at $k_{\rm B}T/\epsilon = 0.77$,
$P\sigma^3/\epsilon = 0.020$, and $x_{\rm B} = 0.78$, which
corresponds to the vapor state.  In these calculations, the left and
right walls of the slit-pore each act as an external potential
confining the particles to a region $-L/2 < z < L/2$
\begin{equation}
  \label{eqn:Vsingle}
  V^{\mrm{(single)}}_{\alpha}(z) =
  4\epsilon_{{\rm w},\alpha}\left[ \left(\sigma/(z-z_{\rm w})\right)^{12} - \left(\sigma/(z-z_{\rm w})\right)^{6} \right],
\end{equation}
where $z_{\rm w}=\pm L/2$. Similar to the interatomic potential, each
$V^{\mrm{(single)}}_{\alpha}$ is truncated and shifted at a cutoff
$z_{\rm c} = 2.5\sigma$. In Fig.~\ref{fig:1}\tcr{b}, we have
considered the symmetric case,
$\epsilon_{{\rm w},\rm A} = \epsilon_{{\rm w},\rm B} = 2.0\,k_{\rm
  B}T$. For a mildly attractive interaction between the confining
walls and the particles, equilibrium densities predicted from \tcr{the
  neural LMFT} calculations are in excellent agreement with GCMC
simulations, and overall consistent with a vapor-like state in the
pore. Upon increasing the interaction strength to $2.5\,k_{\rm B}T$,
\tcr{neural LMFT} captures the transition to a liquid-like state
predicted by GCMC simulations. While some minor discrepancies in the
density profiles are observed for these more attractive walls,
agreement between theory and simulation remains very good.

\begin{figure*}[!htb]
    \centering
    \includegraphics{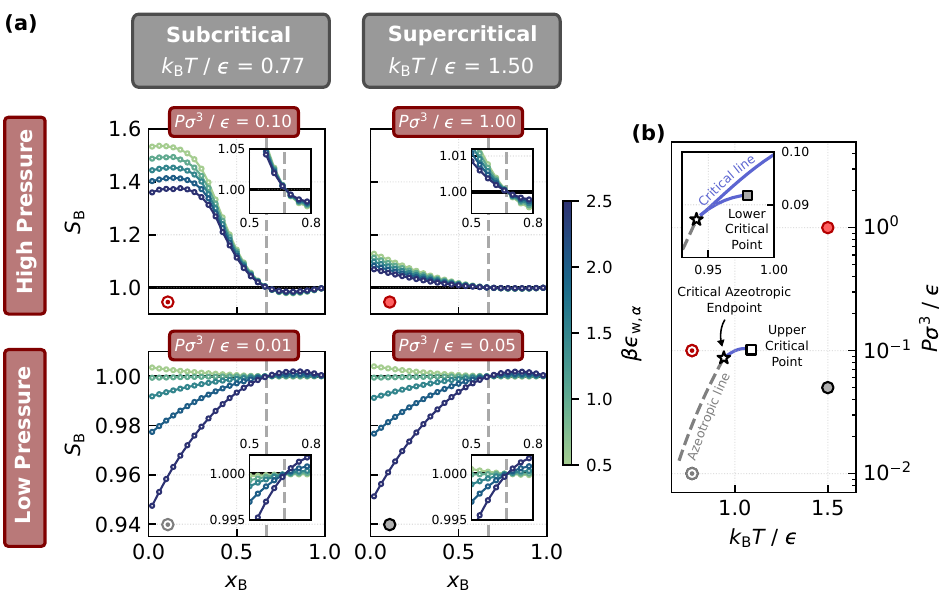}
    \caption{The relevance of the azeotropic composition across a
      broad range of thermodynamic conditions. (a) $S_{\rm B}$ vs
      $x_{\rm B}$ with varying wall--fluid interaction strengths
      \tcr{(indicated by the color bar)} at different state points
      \tcr{(indicated by the labels)}. In all cases, a crossover in
      selectivity occurs at $x_{\rm B}\approx x_{\rm B}^{\rm
        (az)}$. (b) $P$--$T$ phase diagram \tcr{obtained from the PeTS
        EoS}. The critical azeotropic endpoint is
      $k_{\mrm{B}}T_{\text{CAEP}}/\epsilon\approx0.94$
      \cite{staubachInterfacialPropertiesBinary2022} (star) and the
      upper critical point is
      $k_{\mrm{B}}T_{\text{C}}/\epsilon\approx 1.09$ (white
      square). The inset shows the region around
      $k_{\mrm{B}}T_{\text{CAEP}}/\epsilon$ with the lower critical
      point. \tcr{The circles indicate the thermodynamic conditions
        used in (a)}.}  \label{fig:2}
\end{figure*}

Overall, the combination of the accurate PeTS EoS for bulk, and the
neural functional for the inhomogeneous correlations means that
\tcr{the neural LMFT} approach outperforms the standard mean-field
cDFT treatment across a broad range of thermodynamic conditions (see
Figs.~S1--S3). Moreover, we also demonstrate the robustness of our
approach on other LJ binary mixtures with different attractive
interactions \tcr{(we investigate systems with
  $0.85 \le \epsilon_{\rm AB}/\sqrt{\epsilon_{\rm AA}\epsilon_{\rm
      BB}} \le 1.25$)}, illustrating the potential power of the
``train once, learn many'' strategy (see Fig.~S4). \tcr{Clearly,
  success of this approach rests upon the suitability of the reference
  system, and an accurate EoS for the binary fluid. For our current
  purposes, the single-component WCA fluid is reasonable for mixtures
  with components of equal size, and where cross-interactions are not
  too dissimilar, while the PeTS EoS appears sufficiently accurate.}

\subsection{Efficient Evaluation of Pore Selectivity Unveils a
  Significance of the Azeotropic Composition Under Confinement}

Having established the accuracy of \tcr{the neural LMFT} approach, we
now capitalize on its advantages as a density functional
framework. Specifically, we investigate how confinement influences the
overall composition of the fluid across a broad range of wall--fluid
interaction strengths and thermodynamic conditions; this is especially
important for systems that exhibit azeotropy, such as the binary LJ
mixture under investigation here. While thermodynamic modeling of bulk
mixtures is well-established \cite{prausnitz1998molecular}, major gaps
remain in our understanding of azeotropy under confinement
(``adsorption azeotropy''
\cite{doAzeotropicBehaviourAdsorption1999,hamidThermodynamicConditionAdsorption2023,
  ritterAdsorptionBinaryGas2010,
  jiangTheoreticalAnalysisNecessary2021})---addressing this issue is
of broad relevance to chemical separation and industrial processes
\cite{widagdoJournalReviewAzeotropic1996, cholletDesign2021}.

To assess the influence of confinement on the composition of the
fluid, we use \tcr{neural LMFT} to compute the pore selectivity of
each species \cite{jiangTheoryAdsorptionTrace1994,
  qiaoEnhancingGasSolubility2020, huEffectConfinementNanoporous2016,
  coasneGasOversolubilityNanoconfined2019},
\begin{equation}
  \label{eqn:selectivity}
  S_\alpha = \frac{N_\alpha / (N_{\rm A} + N_{\rm B})}{x_\alpha},
\end{equation}
where $N_\alpha$ is the total number of adsorbed particles of species
$\alpha$,
\begin{equation}
N_\alpha=\mathcal{A}\int^{+L/2}_{-L/2}\mrm{d}z\,
\rho_\alpha(z),
\end{equation}
Note that $x_\alpha$ is a property of the bulk reservoir. In the first
instance, we evaluate $S_\alpha$ at fixed $T$ and $P$ over a wide
range of $\epsilon_{\rm w, A} = \epsilon_{\rm w, B}$ and $x_{\rm B}$,
as shown in Fig.~\ref{fig:1}c.

Our calculations of $S_\alpha$ reveal that, relative to the bulk
composition, species B is preferentially adsorbed at low $x_{\rm B}$,
whereas species A is preferred at high $x_{\rm B}$. This observation
is consistent with previous studies that have found selective
adsorption for the component with weaker fluid--fluid interactions when
it is the minority component in bulk
\cite{taghizadehPopulationInversionBinary2011,
  rotherConfinementEffectAdsorption2004}. The most striking
observation from Fig.~\ref{fig:1}c, however, is that the
crossover between B- and A-selectivity, where
$S_{\mrm{A}}=S_{\mrm{B}}=1$, occurs very close to the azeotropic
composition, irrespective of the wall--fluid interaction strength; for
reasons that will become clear in Sec. V, we refer to this
composition, $x^{\mrm{(an)}}$, as the ``aneotropic composition''
\cite{McLure1973, telodagamaStructureSurfaceTension1983, Fouad2017}.

The fact that $x^{\rm (an)}_{\rm B} \approx x_{\rm B}^{\rm (az)}$
appears insensitive to the strength of the wall--fluid interaction
motivates us to investigate the extent to which the azeotropic
composition influences pore selectivity under different thermodynamic
conditions. To this end, we repeat our calculations at the same
temperature $k_{\rm B}T/\epsilon = 0.77$ but at both higher and lower
pressures, and again at $k_{\rm B}T/\epsilon = 1.50$, with the results
shown in Fig.~\ref{fig:2}a.  The state points considered
encompass regions both below and above the azeotropic and critical
lines of the mixture on the $P$--$T$ phase diagram, as marked in
Fig.~\ref{fig:2}b. Remarkably, all show the same crossover at
$x^{\rm (an)}_{\rm B} \approx x_{\rm B}^{\rm (az)}$, despite being far
from the azeotropic line.

So far, we have investigated slit pores that interact with species A
and B in an identical fashion. Yet, even in this simple symmetric
case, we observe selective adsorption. Moreover, we observe completely
unselective behavior close to the azeotropic composition of the
\emph{bulk} fluid. Before going on to investigate how pore selectivity
may vary in the case of asymmetric wall--fluid interactions, we
therefore aim to understand the behavior of the bulk fluid in greater
detail.

\begin{figure*}[!t]
    \centering
    \includegraphics{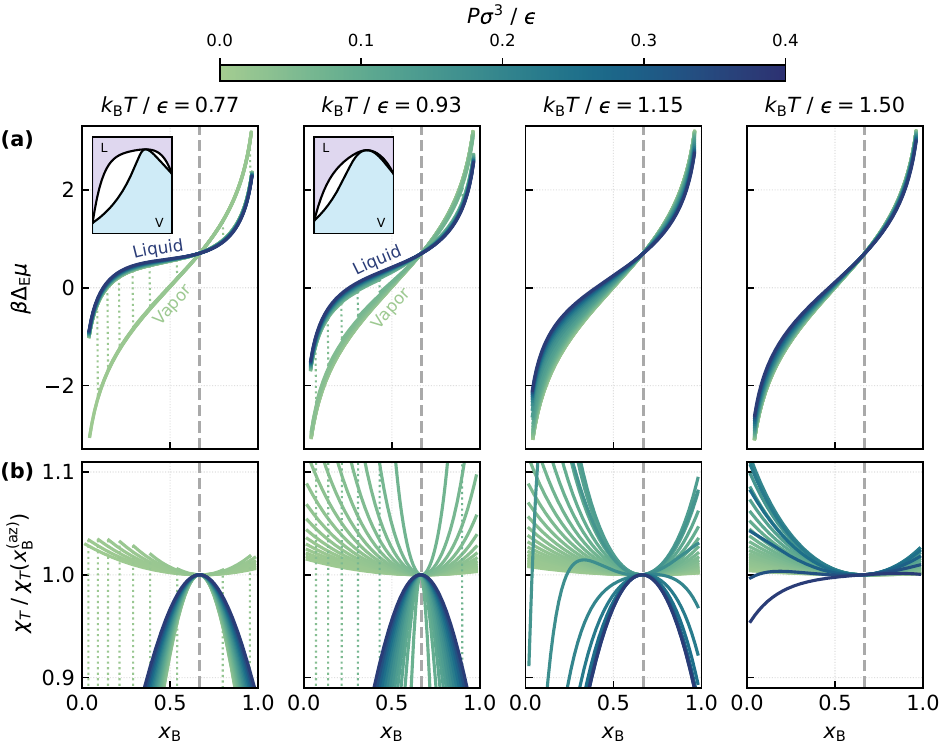}
    \caption{Bulk thermodynamic properties vs $x_{\rm B}$ for
        different $P$ and $T$ \tcr{obtained from the PeTS EoS}. From left to right, the temperatures
        correspond to $T \ll T_{\text{c}}$,
        $T \lesssim T_{\text{CAEP}}<T_{\rm c}$, $T>T_{\text{c}}$, and
        $T \gg T_{\text{c}}$. (a) For subcritical temperatures,
        $\Delta_\text{E} \mu$ collapses into liquid and vapor branches
        that cross at $x_{\rm B}\approx x_{\rm B}^{\rm (az)}$; this
        crossing point persists at supercritical temperatures. The
        insets show schematic representations of the $P$--$x_{\rm B}$
        phase diagram at subcritical temperatures. (b) The isothermal
        compressibility is locally extremum at
        $x_{\rm B}\approx x_{\rm B}^{\rm (az)}$ at all
        temperatures. In both (a) and (b), discontinuities
      representing liquid--vapor phase transitions are shown with
      dotted lines.}  \label{fig:3}
\end{figure*}

\subsection{Bulk Thermodynamics and the Robust Signature of  Azeotropy}

The results we present in Figs.~\ref{fig:1}c and~\ref{fig:2}a indicate
that the azeotropic composition remains relevant to the mixture's
adsorption behavior even at thermodynamic conditions far from
liquid--vapor coexistence. To shed light on these observations, we
recall that the azeotrope is defined as the point of liquid--vapor
coexistence at which \tcr{the} composition of the two phases in bulk
is the same:
$x_{\mrm{B}}^{\rm (l)}=x_{\mrm{B}}^{\rm (v)}=x^{\rm (az)}_{\rm B}$,
where the superscripts ``(l)'' and ``(v)'' indicate quantities that
pertain to the liquid and vapor phases, respectively. Therefore, in
addition to mechanical and thermal equilibrium, we have
\cite{vanKonynenburg1980}
\begin{subequations}
  \begin{align}
    \mu_{\rm B}^{\rm (l)}\big(x^{\rm (az)}_{\rm B},P^{\rm (az)},T^{\rm (az)}\big)= \mu_{\rm B}^{\rm (v)}\big(x^{\rm (az)}_{\rm B},P^{\rm (az)},T^{\rm (az)}\big), \label{eqn:muB-azeo} \\
    \mu_{\rm A}^{\rm (l)}\big(x^{\rm (az)}_{\rm B},P^{\rm (az)},T^{\rm (az)}\big) = \mu_{\rm A}^{\rm (v)}\big(x^{\rm (az)}_{\rm B},P^{\rm (az)},T^{\rm (az)}\big), \label{eqn:muA-azeo}  
  \end{align}
\end{subequations}
where, $T^{\rm (az)}$ and $P^{\rm (az)}$ are the temperature and
pressure along the azeotropic line
\cite{staubachInterfacialPropertiesBinary2022} in the $P$--$T$ plane
(see Fig.~\ref{fig:2}b). Introducing the exchange potential,
\begin{equation}
  \label{eqn:DeltaMu}
  \Delta_{\rm E}\mu(x_{\rm B},P,T) = \mu_{\rm B}(x_{\rm B},P,T) - \mu_{\rm A}(x_{\rm B},P,T),
\end{equation}
and subtracting Eq.~\ref{eqn:muA-azeo} from Eq.~\ref{eqn:muB-azeo}, we
find
\begin{equation}
  \Delta_{\rm E}\mu^{\rm (l)}\big(x^{\rm (az)}_{\rm B},P^{\rm (az)},T^{\rm (az)}\big) =
  \Delta_{\rm E}\mu^{\rm (v)}\big(x^{\rm (az)}_{\rm B},P^{\rm (az)},T^{\rm (az)}\big).
\end{equation}
Thus, under azeotropic conditions, the reversible work required
to exchange a particle of species A for one of species B, as encoded
in $\Delta_{\rm E}\mu$, is identical in the liquid and vapor phases.

Inspired by the observation that the azeotropic composition appears
relevant across a broad range of $P$ and $T$, we investigate the
behavior of the exchange potential away from coexistence. To this end,
in Fig.~\ref{fig:3}a we show how $\Delta_{\rm E}\mu$ varies with
$x_{\rm B}$ for different $P$ and $T$\tcr{; note that these results
  are obtained directly from the PeTS EoS}. At subcritical
temperatures, we see that $\Delta_{\rm E}\mu(x_{\rm B})$ obtained at
different pressures collapse into two distinct branches, corresponding
to the vapor and liquid states. Strikingly, we observe that these two
branches cross at $x_{\rm B}^{\rm (az)}$. At supercritical
temperatures, while no longer separated into liquid and vapor
branches, the observation that
$\Delta_{\rm E}\mu\big(x_{\rm B}^{\rm (az)}\big)$ is equal for all $P$
persists. \tcr{Across the temperature range we consider, these results
  imply that}
\begin{equation}
  \label{eqn:exchange-deriv}
  \left(
\frac{\partial \Delta_{\rm E}\mu}{\partial P}
\right)_{x_{\rm B}=x_{\rm B}^{\rm (az)},T}
= 0.
\end{equation}
An alternative, but equivalent, viewpoint is that at the azeotropic
composition, irrespective of $P$ and $T$, the partial molar volumes,
$\tilde{v}_\alpha = (\partial\mu_{\alpha}/\partial P)_{x,T}$, of A and
B are identical. \tcr{We stress that we have not derived
  Eq.~\ref{eqn:exchange-deriv}; it is an observation based on the
  results in Fig.~\ref{fig:3}a}.

For single component fluids, there has been much interest in
characterizing the nature of the supercritical state. In particular,
while the critical point is defined by the \tcr{indistinguishability} of
liquid and vapor states, several studies suggest regions in the phase
diagram---often separated by so called ``Widom lines''---where the
supercritical state behaves more liquid- or vapor-like
\cite{liThermodynamicCrossoversSupercritical2024, Fomin2015,
  Gorelli2006, Pipich2018, Cockrell2021, Simeoni2010}.  To our
knowledge, the supercritical state of binary fluids remains much less
intensely studied \cite{Raju2017,Saric2022}. In this context, the
observation of identical partial molar volumes at $x_{\rm B}^{\rm
  (az)}$ at $T\gg T_{\rm c}$ is intriguing. While an exhaustive study
into this topic remains beyond the scope of this article, in
Fig.~\ref{fig:3}b we show how the isothermal compressibility,
\begin{equation}
  \chi_T(x_{\rm B},P,T) = \frac{1}{\bar{\rho}}\bigg(\frac{\partial\bar{\rho}}{\partial P}\bigg)_{x_{\rm B},T},
\end{equation}
where $\bar{\rho} = \bar{\rho}_{\rm A} + \bar{\rho}_{\rm B}$, varies
with $x_{\rm B}$ for different $P$ and $T$. Though not unique,
locating conditions of maximum isothermal compressibility is a common
approach to mapping Widom lines
in single-component systems \cite{liThermodynamicCrossoversSupercritical2024,
Fomin2015, Gorelli2006, Pipich2018, Cockrell2021, Simeoni2010}.
For the binary system we consider, we see that $\chi_T\big(x^{\rm
(az)}_{\rm B},P,T\big)$ is locally extremum. Specifically, at
subcritical temperatures, we observe that vapor and liquid states are
characterized by positive and negative curvatures,
respectively. Remarkably, hallmarks of this observation persist at
supercritical temperatures, with $\chi_T\big(x^{\rm (az)}_{\rm B},P,T\big)$
a local minimum at low pressures, and a local maximum at high
pressures. For now, we leave this as an intriguing observation. Our
initial results for pore selectivity (Figs.~\ref{fig:1}c
and~\ref{fig:2}a), however, hint that such supercritical behavior of
the bulk fluid influences behavior under confinement.

\begin{figure}[!tbh]
  \centering
  \includegraphics{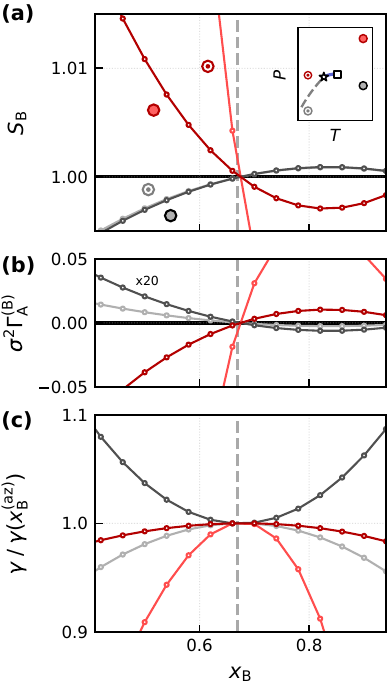}
  \caption{Coincidence of $S_{\rm B}=1$, vanishing relative
    adsorption, and extremal interfacial tension. (a) $S_{\rm B}$ vs
    $x_{\rm B}$ at the four different state points from Fig.~2 (and
    indicated in the inset). (b) $\Gamma_{\rm A}^{\rm (B)}$ vs
    $x_{\rm B}$, with results corresponding to low pressures (light
    and dark grey) multiplied by a factor 20 for clarity. (c)
    \tcr{Wall}--fluid interfacial tension. The vertical dashed line
    indicates the azeotropic composition. All results correspond to a
    slit pore with
    $\beta\epsilon_{\rm w, A} = \beta\epsilon_{\rm w, B} = 2.0$ and
    $L = 8\sigma$.}
  \label{fig:4}
\end{figure}

\subsection{Thermodynamic Origin and Control of Selective Adsorption in Confined Azeotropic Mixtures}
\label{sec:thermodynamic}

\begin{figure*}[!tbh]
    \centering \includegraphics[scale=0.95]{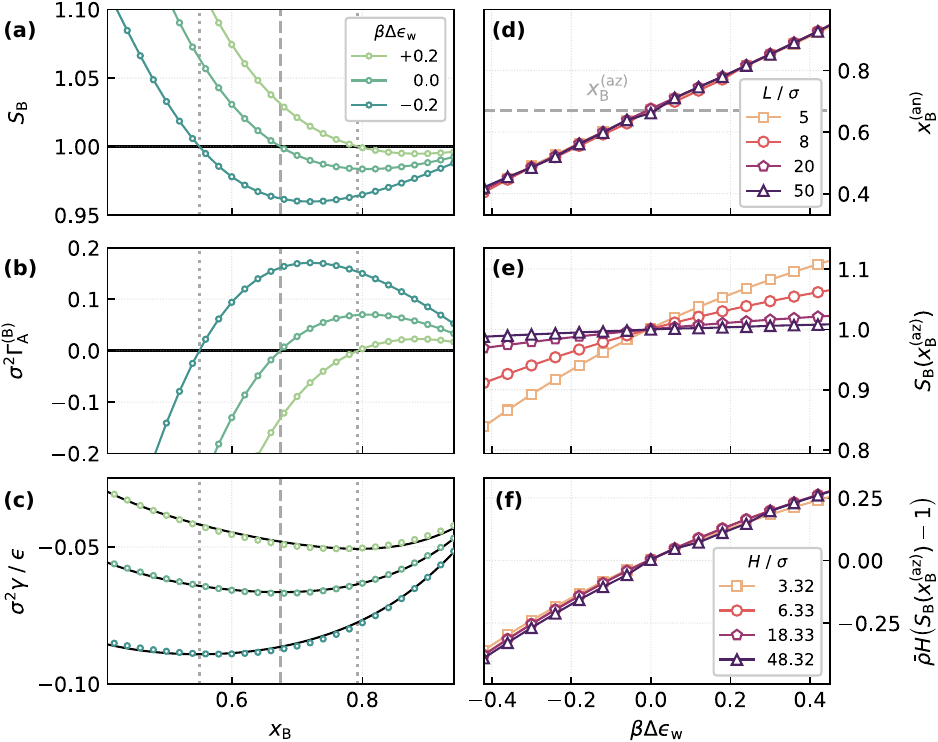}
    \caption{Effect of wall affinity,
      $\Delta\epsilon_{\rm w} = \epsilon_{\rm w,B}-\epsilon_{\rm
          w,A}$, on the pore selectivity. In all cases,
        $\beta \epsilon_{\rm w, B}=2.0$. (a), (b), and (c)
        respectively show how $S_{\rm B}$, $\Gamma_{\rm A}^{\rm (B)}$,
        and $\gamma$ vary with $x_{\rm B}$. In all cases, we observe
        that $S_{\rm B}=1$, $\Gamma_{\rm A}^{\rm (B)} = 0$, and a
        minimum in $\gamma$ coincide. The vertical dashed and dotted
        lines indicate $x_{\rm B}^{\rm (az)}$ and
        $x_{\rm B}^{\rm (an)}$, respectively. The solid black lines in
        (c) show fits to a fourth-order polynomial, constrained to be
        minimum at the aneotropic composition. Panel (d) shows how
        $x_{\rm B}^{\rm (an)}$ varies with
        $\beta\Delta\epsilon_{\rm w}$ for different $L$ (as indicated
        in the legend). (e) $S_{\rm B}(x_{\rm B}^{\rm (az)})$ vs
        $\beta\Delta\epsilon_{\rm w}$ for different widths of the slit
        pore. When rescaled according to Eq.~\ref{eqn:S-by-DeltaEps}
        all data approximately collapse onto the same master curve, as
        seen in (f).}
    \label{fig:5}
\end{figure*}

For the pores that we have considered so far, i.e., those that
interact with species A and B in the same manner, our analysis
indicates that the bulk thermodynamics of the fluid mixture play an
important role in determining selective adsorption under
confinement. While the observations that the partial molar volumes of
A and B are equal at the azeotropic composition, and
$x_{\rm B}^{\rm (an)}\approx x_{\rm B}^{\rm (az)}$, are intriguing,
they do not by themselves provide a clear mechanistic explanation of
the observed adsorption behavior. More generally, there has been
significant interest to understand the mechanisms of selective
adsorption under confinement, especially when B assumes the role of a
minority species dissolved in A, e.g., carbon dioxide in water; in
such cases, $S_{\rm B}$ is often dubbed ``oversolubility.'' In
this context, there has been much work discussing the roles of
adsorption vs confinement-induced changes on the solubility (see,
e.g., Ref.~\onlinecite{coasneGasOversolubilityNanoconfined2019} for a
review). Here, we present a simple thermodynamic analysis that sheds
light on the driving forces that underpin $S_{\rm B}$. Anticipating
the results that follow, we find that, in the thermodynamic sense,
$S_{\rm B}$ is driven solely by adsorption.

Following standard treatments of the thermodynamics of confined fluids
\cite{evansPhaseEquilibriaSolvation1987,
  hansenTheorySimpleLiquids2013}, our starting point is the
exact differential of the surface excess grand potential,
\begin{equation}
  \label{eqn:dOmegaEx}
  \mrm{d}\Omega^{\rm (ex)} = -2s \mcl{A}\,\mrm{d}T - \mcl{A}\sum_\alpha\Gamma_{\alpha}\,\mrm{d}\mu_{\alpha}
  + 2\gamma\,\mrm{d}\mcl{A} - f_{\mrm{s}}\mcl{A}\,\mrm{d}H,
\end{equation}
where $s$ is excess entropy per unit area, $f_{\mrm{s}}$ is the
solvation force, $\gamma$ is the \tcr{wall--fluid} interfacial
tension, \tcr{
\begin{equation}
  \gamma = \frac{\Omega + PV}{2\mcl{A}},
\end{equation}
where $V = \mathcal{A}H$ and $P$ the bulk pressure,}
and
\begin{equation}
 \label{eqn:GammaDef}
 \Gamma_{\alpha} = \int_{-H/2}^{H/2}\!\mrm{d}z\,\big(\rho_{\alpha}(z)-\bar{\rho}_{\alpha}\big)
 = N_\alpha/\mathcal{A} - \bar{\rho}_\alpha H,
\end{equation}
is the adsorption of species $\alpha$. Note that, in this
thermodynamic picture, there is a degree of flexibility in choosing
$H$, the separation between the two walls\tcr{; it need not be the
  same as the separation $L$ that is used in the external potential
  (see Eq.~\ref{eqn:Vsingle}). To be consistent with $N_\alpha$ used
  in Eq.~\ref{eqn:selectivity}, however,} $H$ should be large enough
to encompass all particles in the slit pore \tcr{(we discuss below how
  we choose $H$ in practice)}. As
$\Omega^{\rm (ex)} = 2\gamma\mcl{A}$, from Eq.~\ref{eqn:dOmegaEx} one
can straightforwardly obtain the Gibbs adsorption equation for
mixtures,
\begin{equation}
  2\,\mrm{d}\gamma + 2s\,\mrm{d}T + \sum_{\alpha} \Gamma_{\alpha}\,\mrm{d}\mu_{\alpha} + f_{\mrm{s}}\,\mrm{d}H = 0,
\end{equation}
from which it immediately follows that
\begin{equation}
  \label{eqn:GammaA-gammaDeriv}
  \Gamma_{\alpha} =
  -2\left(\partial\gamma /\partial\mu_\alpha\right)_{\mu_{\eta \neq \alpha},H,T}.
\end{equation}

From the definitions of $S_{\alpha}$ (Eq.~\ref{eqn:selectivity}) and
$\Gamma_\alpha$ (Eq.~\ref{eqn:GammaDef}), we obtain
\begin{equation}
  \label{eqn:SB-GammaA}
  S_{\rm B} = \frac{1}{x_{\rm B}}\frac{\Gamma_{\rm B}+\bar{\rho}_{\rm B}H}{\Gamma_{\rm A} + \Gamma_{\rm B} + (\bar{\rho}_{\rm A}+\bar{\rho}_{\rm B})H},
\end{equation}
This exact relation makes explicit that the adsorptions of species A
and B are the key thermodynamic quantities governing the pore
selectivity. Note that $S_{\rm B}$ is independent of \tcr{the}
definition of $H$.  Given the flexibility afforded by this invariance,
we choose $H$ such that $\Gamma_{\rm B} = 0$ when $S_{\rm B} = 1$;
this is only possible if $\Gamma_{\rm A}$ also vanishes. That is, the
relative adsorption,
\begin{equation}
  \Gamma_{\rm A}^{\rm (B)} = \Gamma_{\rm A} - (\bar{\rho}_{\rm A}/\bar{\rho}_{\rm B})\Gamma_{\rm B},
\end{equation}
is zero at $S_{\rm B} = 1$. This result is confirmed in Figs.~4a
and~4b, where we show how $S_{\rm B}$ and $\Gamma_{\rm A}^{\rm (B)}$,
respectively, vary with $x_{\rm B}$ at various different thermodynamic
state points, for a slit pore with
$\beta\epsilon_{\rm w, A} = \beta\epsilon_{\rm w, B} = 2.0$ and
$L = 8\sigma$.

The point at which the relative adsorption is zero defines the
aneotropic composition \cite{McLure1973,
  telodagamaStructureSurfaceTension1983, Fouad2017}. For the
liquid--vapor interface of a similar binary LJ mixture to that studied
here, Telo da Gama and Evans
\cite{telodagamaStructureSurfaceTension1983} also reported that
$x^{\rm (an)}_{\rm B}\approx x^{\rm (az)}_{\rm B}$, for which they
provided a qualitative explanation: at low $x_{\rm B}$, the vapor
phase is relatively enriched in species B, leading to preferential
adsorption of B; at $x_{\rm B} = x_{\rm B}^{\rm (az)}$, the liquid and
vapor phases have identical compositions and the adsorption vanishes;
for high $x_{\rm B}$ the liquid becomes relatively richer in species B
than the vapor, causing the adsorption to change sign. In other words,
the sign of $\Gamma^{\rm (B)}_{\rm A}$ is controlled by whether the
bulk vapor or bulk liquid phase is enriched in species B. \tcr{Based
  on our results, it would seem that, when the walls interact with A
  and B in the same manner, a similar qualitative understanding
  extends to fluids under confinement.}

Telo da Gama and Evans also found that the liquid--vapor surface
tension is minimum at the aneotropic point. \tcr{It is natural, then,
  to explore the extent to which the aneotropic point influences the
  wall--fluid interfacial tension in these confined systems.} From
Eq.~\ref{eqn:GammaA-gammaDeriv}, and using the Gibbs--Duhem relation,
we find
\begin{equation}
    \left(\frac{\partial\gamma}{\partial x_{\mathrm{B}}}\right)_{H, P, T} =
-\frac{1}{2}\left(\frac{\partial\mu_{\rm A}}{\partial x_{\rm B}}\right)_{P, T}\bigg[\Gamma_{\rm A} - \bigg(\frac{1-x_{\rm B}}{x_{\rm B}}\bigg)\Gamma_{\rm B}\bigg].
\end{equation}
Evaluating at $x_{\rm B} = x_{\rm B}^{\rm (an)}$, where
$\Gamma_{\rm A} = \Gamma_{\rm B} = 0$, we have
\begin{equation}
  \label{eqn:extremum-surfacetension}  
  \left(\frac{\partial\gamma}{\partial x_{\mrm{B}}}\right)_{H, P, T}\bigg|_{x_{\mrm{B}}= x^{\mrm{(an)}}_{\mrm{B}}} = 0.
\end{equation}
While this result is consistent with the minimum \tcr{of the
  liquid--vapor surface tension} reported in
Ref.~\onlinecite{telodagamaStructureSurfaceTension1983}, more
generally, it states that the aneotropic point corresponds to \tcr{an
  extremum in wall-fluid interfacial tension.} This prediction is
borne out by results from \tcr{neural LMFT}; as seen in Fig.~\ref{fig:4}c,
$\gamma(x^{\rm (an)})$ changes from a local minimum to a local maximum
as the system traverses from high to low pressures.

The above arguments readily extend to cases where the relative wall
affinity
$\Delta\epsilon_{\rm w} = \epsilon_{\rm w, B} - \epsilon_{\rm w, A}
\neq 0$. Intuitively, we expect that \tcr{varying}
$\Delta\epsilon_{\rm w}$ will change the pore selectivity, and the
aneotropic point. This notion is confirmed in Figs.~5a, 5b, and~5c,
where we respectively show how $S_{\rm B}$,
$\Gamma_{\rm A}^{\rm (B)}$, and $\gamma$ vary with $x_{\rm B}$, with
$\beta\Delta\epsilon_{\rm w} = \pm 0.2$, for a system at
$k_{\rm B}T/\epsilon = 0.77$, $P\sigma^3/\epsilon = 0.0248$, and
$L = 8\sigma$. As expected, we see that $S_{\rm B} = 1$ \tcr{coincides
  with} the aneotropic composition. The results also appear consistent
with a minimum in interfacial tension, albeit a shallow one. Due to
this shallowness, we have performed fits to a fourth-order polynomial,
constrained to be minimum at $x_{\rm B}^{\rm (an)}$, as shown by the
solid black lines in Fig.~\ref{fig:5}c.

In Fig.~\ref{fig:5}d, we show how $x^{\rm (an)}$ varies with
$\beta\epsilon_{\rm w}$ for different slit widths. To a very good
approximation, the variation is linear, and insensitive to $L$. We
interrogate this size-independence further by considering how
selectivity varies with $\Delta\epsilon_{\rm w}$ within the context of
the thermodynamic model presented above. As we have established that
$x_{\rm B}^{\rm (an)}\approx x_{\rm B}^{\rm (az)}$ when
$\Delta\epsilon_{\rm w} = 0$, it is a straightforward, though slightly
tedious, matter to show that
\begin{equation}
  \label{eqn:S-by-DeltaEps}
  \bar{\rho}H\big(S_{\rm B}(x^{\rm (az)}_{\rm B};\Delta\epsilon_{\rm w}) - 1\big) =
  -\Delta\epsilon_{\rm w}\left(\frac{\partial\Gamma^{(\rm B)}_{\mathrm{A}}(x_{\rm B}^{\rm (az)};0)}{\partial\Delta\epsilon_{\rm w}}\right)_{H,P,T}.
\end{equation}
In Fig.~\ref{fig:5}e we show how $S_{\rm B}(x^{\rm (az)}_{\rm B})$ varies with
$\Delta\epsilon_{\rm w}$, where we see that \tcr{the} variation is
nonlinear, which becomes more pronounced for smaller slit pores. In
Fig.~\ref{fig:5}f, we plot the left hand side of Eq.~\ref{eqn:S-by-DeltaEps} vs
$\Delta\epsilon_{\rm w}$ for different slit widths. Strikingly, we
observe that all data approximately collapse onto the same curve. This
insensitivity to the slit width suggests that, even for systems barely
large enough to accommodate three layers of particles, the two
interfaces of the slit pore essentially behave independently; in this
case, down to the range of the wall--fluid interaction.

\section{Conclusions}

In this study, we have extended the recently developed neural density
functional theory to a binary mixture of Lennard--Jones particles that
exhibits azeotropic phase behavior. In contrast to another recent
neural functional theory study on mixtures
\cite{robitschkoLearningBulkInterfacial2025}, we have used machine
learning to obtain an accurate representation of a \emph{single
  component} repulsive reference system, and treated attractive
interactions in a mean-field fashion. \tcr{The mean-field approach
  that we adopt is rooted in the connection between classical density
  functional theory (and its recent extension, hyperdensity functional
  theory), and \tcr{local molecular field theory} derived by Weeks and
  co-workers} \cite{weeksRolesRepulsiveAttractive1998,
    weeksSelfConsistentTreatmentRepulsive1995,
    remsingLongrangedContributionsSolvation2016}. Within this
\tcr{neural LMFT} framework, we have taken advantage of the fact that,
when known from other sources
\cite{heierEquationStateLennardJones2018,
  rehnerFeOsOpenSourceFramework2023}, the bulk equation of state of
the fluid can be integrated seamlessly, allowing us to focus \tcr{on}
applying density functional theory itself to inhomogeneous systems.

We have used this \tcr{neural LMFT} framework to understand
preferential adsorption of this binary fluid in a slit-pore
geometry. In cases where the walls of the slit pore interact with both
species of the fluid in the same manner, our numerical results
indicate that when the reservoir is at its azeotropic composition, so
too is the composition in the pore. Remarkably, this observation
persists across a broad range of thermodynamic conditions, including
far into the supercritical state. By analyzing the bulk equation of
state, we find that the azeotropic composition coincides with equal
partial molar \tcr{volumes} of the constituent species, and a local
extremum of the isothermal compressibility. Intriguing as these
observations are, their generality to other fluids that exhibit
azeotropy is an open question that warrants further investigation. For
example, the system we have investigated has an azeotropic composition
that is largely \tcr{insensitive to changes in} temperature and
pressure; this does not generally hold, especially when the components
have markedly different sizes
\cite{panditEvaluationLocusAzeotropes1999}.

To elucidate the mechanisms underlying pore selectivity, we have
presented a thermodynamic description that connects selectivity
directly to interfacial adsorption.  This shows that the aneotropic
composition, defined by vanishing relative adsorption, remains closely
tied to the bulk azeotropic composition over a wide range of
thermodynamic conditions, even under confinement.  We show clearly
that the relative adsorption of each species in the mixture is the
relevant thermodynamic driving force, with the wall--fluid interfacial
tension reaching an extremum at the aneotropic composition---a result
that generalizes previous work on the \tcr{liquid--vapor interface}
\cite{telodagamaStructureSurfaceTension1983} \tcr{to confined
  systems}.  By analyzing our numerical results within the context of
this thermodynamic model, we found that the aneotropic composition
shifts linearly with the relative affinity of the confining walls to
the two species.  Moreover, we found that the two interfaces of the
slit pore act essentially independently down to remarkably small
separations---a little over three molecular diameters---between the
walls.

With machine learning techniques, it is now possible to accurately
model systems of remarkable complexity---both in terms of their
interactions and emergent phase behavior---using classical density
functional theory. Here, we take ``classical density functional
theory'' in a broad sense to encompass its recent extensions hyper-DFT
\cite{sammullerHyperdensityFunctionalTheory2024} and meta-DFT
\cite{kampaMetadensityFunctionalTheory2025}. \tcr{We} have used
\tcr{the connections between local molecular field theory and cDFT} as
a means to justify a mean-field treatment of attractive interactions,
as well as to incorporate an established bulk equation of state into
the framework. By learning the one-body direct correlation function
once for a single-component system, which we then used as reference
for a binary fluid, we have demonstrated a ``train once, learn many''
strategy as a proof-of-principle. For more complex systems, especially
mixtures where particle sizes differ significantly, \tcr{it may prove}
fruitful to combine this strategy with meta-DFT, which aims to learn
the functional dependence on the interaction potential
directly. \tcr{Other developments in applying ML to obtain bulk
  equations of state may also prove useful}
\cite{chaparro2023development,chaparro2024development}.  Irrespective
of the exact strategy that one adopts, it seems highly likely that
classical density functional theory combined with machine learning
will play an increasingly important role in understanding fluids
relevant to physical chemistry.

\section*{Acknowledgements}
S.J.C. is Royal Society University Research Fellow at Durham
University (Grant No. URF\textbackslash R1\textbackslash
211144). A.T.B. acknowledges funding from the Oppenheimer Fund and Peterhouse College,
University of Cambridge. This work made use
of the Hamilton HPC Service of Durham University and facilities of
the N8 Centre of Excellence in Computationally Intensive Research (N8
CIR) provided and funded by the N8 research partnership and EPSRC
(Grant No. EP/T022161/1). The Centre is coordinated by the
Universities of Durham, Manchester, and York.

\section*{Data Availability}

Data and code supporting the findings of this study are openly
  available at \tcr{Zenodo\cite{data} and Github (\url{https://github.com/CoxGroup/lj-cdft-lmft})}.

\section*{Supporting information}

Supporting information includes additional results validating the
accuracy of the mean-field neural LMFT approach, comparison to
standard mean-field cDFT, and additional calculations for other binary
LJ mixtures.

\bibliography{reference}

\end{document}

\title{Supplementary Information: Roles of Bulk and Surface
  Thermodynamics in the Selective Adsorption of a Confined Azeotropic
  Mixture}

\author{Katie L. Y. Zhou}

\affiliation{Department of Chemistry, Durham University, South Road, Durham, DH1
3LE, United Kingdom}

\author{Anna T. Bui}

\affiliation{Yusuf Hamied Department of Chemistry, University of
  Cambridge, Lensfield Road, Cambridge, CB2 1EW, United Kingdom}
\affiliation{Department of Chemistry, Durham University, South Road, Durham, DH1
3LE, United Kingdom}

\author{Stephen J. Cox}
\email{stephen.j.cox@durham.ac.uk}

\affiliation{Department of Chemistry, Durham University, South Road, Durham, DH1
3LE, United Kingdom}

\date{\today}

\maketitle
\thispagestyle{plain}

\section{Comparison to Simulation and Standard Mean-Field Theory} 

Archer and Evans previously established a connection between
LMFT and traditional mean-field cDFT
\cite{archerRelationshipLocalMolecular2013}. While similar to our
formalism, $\Delta\mu$ was identified as an integration constant,
which for mixtures reads
\begin{equation}
  \label{eqn:mf-mu}
  \Delta\mu_{\alpha} =
  \sum_{\eta} \bar{\rho}_{\eta}
  \int\!\mrm{d}\mbf{r}^\prime\,u_{1,\alpha\eta}(|\mbf{r}^\prime|) \;.
\end{equation}
In this section, we compare the neural LMFT approach to the
standard mean-field cDFT treatment. That is, we consider the above
expression (Eq.~\ref{eqn:mf-mu}) for $\{\Delta\mu_\alpha\}$ together
with a hard sphere (HS) reference; here we use the accurate neural
functional from Ref.~\onlinecite{sammullerNeuralFunctionalTheory2023}.

\begin{figure}
    \centering
    \includegraphics{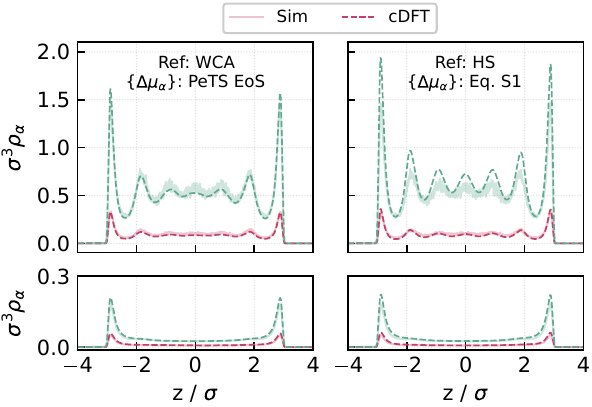}
    \caption{Comparison with the standard mean-field cDFT treatment
      for capillary condensation. Left: neural LMFT, using a WCA
      reference and $\{\Delta\mu_\alpha\}$ from the PeTS equation of
      state. Right: standard mean-field cDFT, using a HS reference and
      Eq.~\ref{eqn:mf-mu} for $\{\Delta\mu_\alpha\}$. Top row shows
      the condensed liquid state with
      $\epsilon_{{\rm w},\rm A} = \epsilon_{{\rm w},\rm B} =
      2.5\,k_{\rm B}T$, while the bottom row shows the vapor state at
      $\epsilon_{{\rm w},\rm A} = \epsilon_{{\rm w},\rm B} =
      2.0\,k_{\rm B}T$.}
    \label{fig:si_cap-cond}
\end{figure}

In Fig. \ref{fig:si_cap-cond} we show a comparison between the
  neural LMFT framework and standard mean-field cDFT for capillary
condensation (Fig. 1b in the main text), with a reservoir at
$k_{\rm B}T/\epsilon = 0.77$, $P\sigma^3/\epsilon = 0.020$, and
$x_{\rm B} = 0.78$, which corresponds to the vapor state. Both
approaches show the transition from a gas-like to liquid-like state
when going from interaction strength
$\epsilon_{{\rm w},\rm A} = \epsilon_{{\rm w},\rm B} = 2.0 k_{\rm B}T$
to $2.5 k_{\rm B}T$. While both are in excellent agreement with GCMC
simulation for the gas-like state, they begin to deviate for the
higher-density liquid-like state. Minor discrepancies are observed
with neural LMFT, while the standard mean-field result
overestimates the peaks in the density profiles.

\begin{figure}
    \centering
    \includegraphics{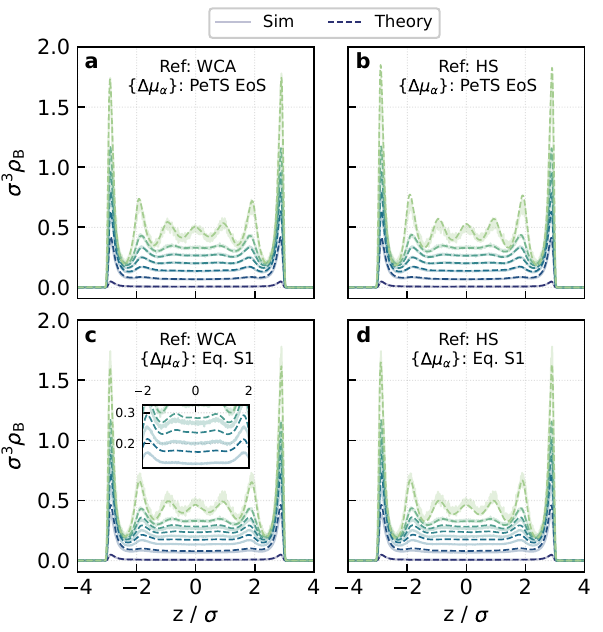}
    \caption{Comparison of reference systems and cDFT approaches at
      different bulk densities at supercritical conditions. (a)
      Neural LMFT, using the WCA reference and the PeTS equation
      of state. (b) HS reference and the PeTS equation of state. (c)
      WCA reference and $\{\Delta\mu_\alpha\}$ given by
      Eq.~\ref{eqn:mf-mu}. (d) Standard mean-field cDFT, using a HS
      reference and $\{\Delta\mu_\alpha\}$ given by
      Eq.~\ref{eqn:mf-mu}.}
    \label{fig:si_cdft-vs-mft}
\end{figure}

In Fig.~\ref{fig:si_cdft-vs-mft} we investigate the factors governing
these differences in more detail, such as the reference system and the
treatment of $\{\Delta\mu_\alpha\}$. The reservoir is now in a
supercritical state at $k_{\rm B}T/\epsilon = 1.50$ and
$x_{\rm B} = 0.66$, with the total density varied between
$\sigma^3\bar{\rho} = 0.01$ to $\sigma^3\bar{\rho} = 0.7$. The two walls are separated
by $L = 8 \sigma$ and interact with strength
$\epsilon_{{\rm w},\rm A} = \epsilon_{{\rm w},\rm B} = 2.0\,k_{\rm
  B}T$. As observed in Fig.~\ref{fig:si_cdft-vs-mft}a, neural
  LMFT, i.e., using the WCA reference and the PeTS equation of state
for $\{\Delta\mu_\alpha\}$, results in excellent agreement with
GCMC simulation across all densities. Switching to a HS reference in
Fig.~\ref{fig:si_cdft-vs-mft}b results in similarly excellent
agreement for most densities, though the system is over-structured at
$\sigma^3\bar{\rho} = 0.7$. In Figs.~\ref{fig:si_cdft-vs-mft}c
and~\ref{fig:si_cdft-vs-mft}d, we respectively use WCA and HS as
reference, but $\{\Delta\mu_\alpha\}$ is obtained with
Eq.~\ref{eqn:mf-mu}. While both reference systems show excellent
agreement at the highest and lowest densities, neither obtains the
correct bulk density at intermediate densities.  This is in line with
observations made with LMFT
\cite{weeksIntermolecularForcesStructure1997,
  weeksRolesRepulsiveAttractive1998}.

\begin{figure}
    \centering
    \includegraphics{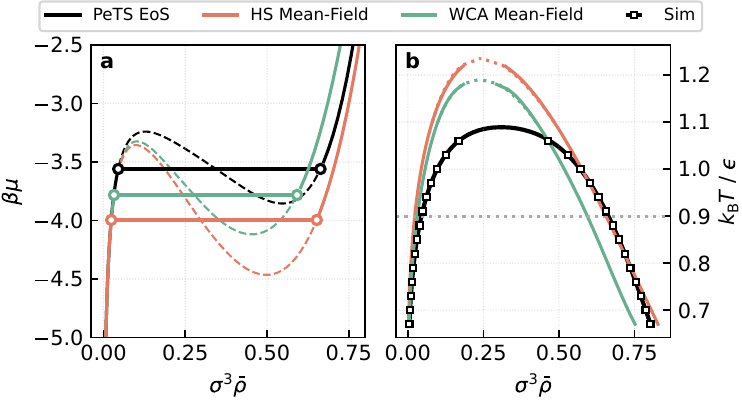}
    \caption{Bulk liquid--vapor coexistence for the single-component
      LJ fluid, comparing the EoS against the standard mean-field
      treatment with different reference systems. (a) The chemical
      potential as a function of bulk density at
      $k_{\rm B}T/\epsilon = 0.90$, highlighted by the dotted grey
      line on the right. The coexistence densities obtained from
      Maxwell construction are shown by the circles. (b) The binodal
      obtained from Maxwell construction. Fits of the binodal to
      Eq.~\ref{eq:binodal_fit} are shown by the colored dotted
      lines. Simulation results are from Vrabec \etal{} in
      Ref.~\onlinecite{vrabecComprehensiveStudyVapour2006}.}
    \label{fig:si_binodal}
\end{figure}

Highlighting further the differences in bulk thermodynamics used by
mean-field theory and the two reference systems, we present the
binodal in Fig.~\ref{fig:si_binodal} for the single-component
truncated and shifted LJ fluid. The neural LMFT approach, by
construction, reproduces the same outputs as that from the PeTS
equation of state. Fig.~\ref{fig:si_binodal}a shows an isotherm of the
chemical potential as a function of bulk density at
$k_{\rm B}T/\epsilon = 0.90$. All approaches show a van der Waals
loop. Performing a Maxwell construction on these isotherms gives the
binodal in Fig.~\ref{fig:si_binodal}b. Comparing to simulation results
from Ref.~\onlinecite{vrabecComprehensiveStudyVapour2006}, the PeTS
equation of state (and by extension, neural LMFT) is the most
accurate. The mean-field theories can accurately reproduce the vapor
coexistence densities, though the WCA reference underestimates
the liquid densities. The HS reference, through cancellation of
errors, results in accurate liquid densities below the critical
temperature.

In addition to the errors with the coexistence densities, the
mean-field approaches overestimate the critical temperature and
underestimate the critical density. Near the critical temperature, an
empirical fit can be made of the form
\cite{sammullerNeuralDensityFunctional2025,
  wildingCriticalpointCoexistencecurveProperties1995}
\begin{equation} \label{eq:binodal_fit}
  \rho_{\pm} = a|T - T_{\text{c}}| \pm b|T - T_{\text{c}}|^\beta + \rho_{\text{c}}
\end{equation}
where $\rho_+ = \rho^{\rm (l)}$ and $\rho_- = \rho^{\rm (v)}$ are the liquid and vapor
coexistence densities respectively, $\rho_{\text{c}}$ is the critical
density, $\beta$ is the standard 3D Ising result
\cite{ferrenbergPushingLimitsMonte2018}, and $a$ and $b$ are
parameters to be fitted. Results are presented in Table
\ref{tab:binodal_fit}. The PeTS equation of state returns values that
are closest to those observed in literature. 

\begin{table}[!tb]
\centering
\caption{Critical exponents extracted from the fits to the binodals (Eq. \ref{eq:binodal_fit}). The number in parentheses indicates the uncertainty in the last digit.}
\label{tab:binodal_fit}
\begin{tabular}{c c c c c}
\toprule
\midrule
{}  & HS Mean-Field & WCA Mean-Field & PeTS & Literature \\
\midrule
$k_{\rm B}T_{\text{c}}/\epsilon $ & $1.236(1)$ & $1.1909(6) $& $1.089$ \cite{stephanVaporLiquidInterfaceLennardJones2018}  & $1.0779$ \cite{vrabecComprehensiveStudyVapour2006} \\
$\sigma^3\rho_{\text{c}}$ & $0.2422(7)$ & $0.2381(4) $& $0.3092$ \cite{stephanVaporLiquidInterfaceLennardJones2018} &  $0.3190$ \cite{vrabecComprehensiveStudyVapour2006} \\
$\beta$ & $0.500(3)$ & $0.472(2)$ & $0.350(2)$ & $0.32630(22)$ \cite{ferrenbergPushingLimitsMonte2018} \\
\midrule
\bottomrule
\end{tabular}
\end{table}

\newpage

\section{Results For Other Systems} 

\begin{figure}[h]
    \centering
    \includegraphics{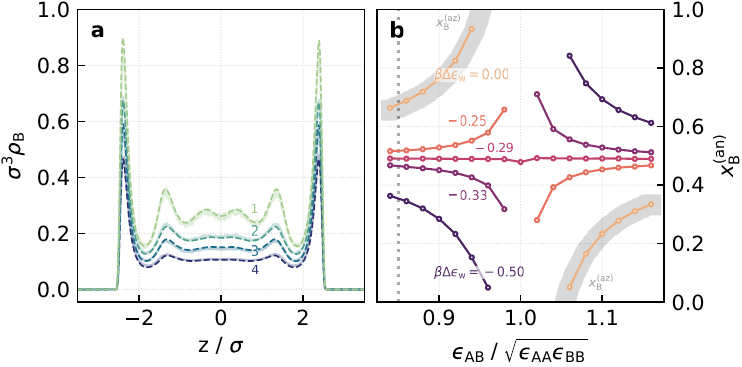}
    \caption{Varying $\epsilon_{\rm BB}$ and $\epsilon_{\rm AB}$ of
      binary LJ mixtures. (a) Density profiles of different binary LJ
      systems; the parameters are summarized in Table
      \ref{tab:mixture-param}. Excellent agreement to GCMC simulations
      can be seen. (b) The position of the aneotrope
      ($S_{\text{B}} = 1$) calculated for varying
      $\Delta \epsilon_{{\rm w}}$ and $\epsilon_{\rm AB}$ for
      systems with $\epsilon_{\rm BB} = 0.9 \epsilon$. The shaded grey
      regions indicate the azeotropic composition, obtained from the
      PeTS equation of state \cite{heierEquationStateLennardJones2018,
        rehnerFeOsOpenSourceFramework2023}. The dotted line indicates
      the system studied in the main paper i.e.,
      $\epsilon_{\rm AB} =0.806\epsilon$. }
    \label{fig:other_mixtures}
\end{figure}

One of the main benefits of using a single-component reference system
is that it lends itself readily to a ``train once, learn many''
strategy. As further demonstration of this transferability, we show
brief results in Fig.~\ref{fig:other_mixtures}. Density profiles of
truncated and shifted LJ binary mixtures with varying
$\epsilon_{\rm BB}$ and $\epsilon_{\rm AB}$ are presented in
Fig.~\ref{fig:other_mixtures}a and compared to GCMC
simulation. Parameters of the mixtures are summarized in Table
\ref{tab:mixture-param}. The bulk reservoir is at a supercritical
state with $k_{\rm B}T/\epsilon = 1.50$, with
$\beta\mu_{\rm A} = \beta\mu_{\rm B} = -2.0$. The two walls interact
with strength
$\epsilon_{{\rm w},\rm A} = \epsilon_{{\rm w},\rm B} = 2.0\,k_{\rm
  B}T$. Excellent agreement between neural LMFT and the
simulation results can be seen. 
The same pore selectivity analysis conducted in the main paper can easily be extended to these other mixtures. In Fig.~\ref{fig:other_mixtures}b, results are
presented for the position of the aneotrope with
$\epsilon_{{\rm w},\rm B} = 2.0\,k_{\rm B}T$ and varying
$\Delta \epsilon_{{\rm w}}$ for systems confined in a slit of width
$L = 8 \sigma$. The bulk reservoir is at $k_{\rm B}T/\epsilon = 0.77$
and $P\sigma^3/\epsilon = 0.10$, a bulk liquid for all systems. The
interaction between unlike species is varied whilst
$\epsilon_{\rm BB}$ is kept at $0.9\epsilon$. Like with the system
studied in the main paper, we find that for the case of symmetric
walls $x^{\rm (an)}_{\rm B} \approx x_{\rm B}^{\rm (az)}$. Varying the
asymmetry of the wall potential leads to shifts in
$x^{\rm (an)}_{\rm B}$.

\begin{table}[h]
\centering
\caption{Parameters of the LJ binary mixtures in Fig.~\ref{fig:other_mixtures}a. Mixture 3 is the same as that studied in the main paper.}
\label{tab:mixture-param}
\begin{tabular}{c c c}
\toprule
\midrule
Mixture & $\epsilon_{\rm BB} / \epsilon_{\rm AA}$ & $\epsilon_{\rm AB} / \sqrt{\epsilon_{\rm AA}\epsilon_{\rm BB}}$ \\
\midrule
1 & 0.5 & 0.85 \\
2 & 0.5 & 1.25 \\
3 & 0.9 & 0.85 \\
4 & 0.9 & 1.25 \\
\midrule
\bottomrule
\end{tabular}
\end{table}

%----------------------------------------------------------------------
\bibliography{reference}
%----------------------------------------------------------------------